\documentclass[superscriptaddress,aps,preprintnumbers,amsmath,amssymb,prd,nofootinbib,reprint]{revtex4-1}
\pdfoutput=1
\usepackage{booktabs} 
\usepackage{array} 
\usepackage[table,xcdraw]{xcolor} 
\usepackage{amsmath,amssymb}
\usepackage{graphicx}
\usepackage{fancyhdr}
\usepackage{bm, color}
\usepackage{slashed,amsthm,amsfonts,empheq}
\usepackage[caption=false]{subfig}
\usepackage{hyperref}
\hypersetup{hidelinks}
\usepackage{booktabs}
\usepackage{multirow}
\usepackage[left=2cm,right=2cm,top=2cm,bottom=2cm,includefoot,a4paper]{geometry}
\usepackage{tikz}
\usepackage{placeins}
\usetikzlibrary{arrows.meta}
\usepackage{amsmath}

\newcommand{\Slash}[1]{{\ooalign{\hfil#1\hfil\crcr\raise.167ex\hbox{/}}}}

\newcommand{\beq}{\begin{equation}}  \newcommand{\eeq}{\end{equation}}
\newcommand{\bef}{\begin{figure}}  \newcommand{\eef}{\end{figure}}
\newcommand{\bec}{\begin{center}}  \newcommand{\eec}{\end{center}}

\newcommand{\laq}[1]{\label{eq:#1}}  
\newcommand{\Eq}[1]{Eq.~(\ref{eq:#1})}

\newcommand{\vev}[1]{\left\langle {#1} \right\rangle}

\def\({\left(}
\def\){\right)}

\def\O{\mathcal{O}}

\newcommand{\GEV}{\,{\rm GeV}}
\newcommand{\TEV}{\,{\rm TeV}}

\def\l{\lambda}

\def\*{\dagger}

\usepackage{amsmath,amssymb,bm}
\usepackage{hyperref}
\hypersetup{hidelinks}

\newcommand{\mpl}{M_{\rm Pl}}

\def\WY#1{#1}

\begin{document}

\title{\WY{A PQ-Symmetric High-Scale SUSY Interpretation of the LZ High-Energy Recoil}}

\author{Wen Yin}
\affiliation{Department of Physics, Tokyo Metropolitan University,
Minami-Osawa, Hachioji-shi, Tokyo 192-0397, Japan}

\begin{abstract}
The 248-keV nuclear-recoil-like event reported by LZ can be interpreted as near-threshold upscattering of 1.08-TeV thermal Higgsino dark
matter.  A neutral-Higgsino splitting of order $0.3$--$0.35$ MeV places the
fixed weak-interaction cross section close to the observed LZ interval and
implies electroweak gauginos at ${\cal O}(10^7)$ GeV.  I point out that this spectrum is
compatible with gravity-mediated high-scale supersymmetry and a Kim--Nilles
Peccei--Quinn (PQ) sector.  
Although the GUT scale is predicted to be lower than the conventional one, proton decay is suppressed due to the high SUSY  scale which also solves the conventional CP, flavor, gravitino and moduli problems. Indeed, proton decay may be probed in the future. 
The PQ symmetry lowers the Higgsino mass below the SUSY scale and
also yields a QCD axion that solves the strong-$CP$ problem but with a cancelled photon coupling due to the Higgsino contribution. If the axion and Higgsino
constitute mixed dark matter, the halo-profile-dependent H.E.S.S. limit can be  relaxed. 
\end{abstract}

\maketitle

\section{Inelastic Higgsino and high-scale supersymmetry}

LZ reports one event consistent with a
$248\pm23_{\rm stat}\pm23_{\rm sys}$ keV nuclear recoil \WY{in a
very-low-background region}.  The maximum local and global significances are $3.4\sigma$
and $2.6\sigma$, respectively, and \WY{the analysis is non-blind; I therefore
use the event only as a working hypothesis}~\cite{LZ:2026highE}.
\WY{Although this is only one event, it points to an interesting possibility
that I explore below.}

A nearly pure Higgsino is a pseudo-Dirac
fermion~\cite{Nagata:2014wma,Graham:2024syw}.  Its $Z$ coupling is
off-diagonal, so direct detection proceeds through
$\widetilde H_1 N\to\widetilde H_2 N$.  Since the momentum transfer of the
event, $q\simeq246$ MeV, is much smaller than $m_Z$, this is a contact
interaction. 
 In particular,
\begin{equation}
 v_{\min}(E_R)=\frac{m_AE_R/\mu_A+\delta_0}
 {\sqrt{2m_AE_R}},\qquad
 E_R^\star=\delta_0\frac{\mu_A}{m_A}.
\end{equation}
For a 1.08-TeV Higgsino and xenon, $\mu_A/m_A\simeq0.90$; hence the recoil
location alone points to $\delta_0\simeq0.28$ MeV.  The normalization is fixed,
\begin{align}
 \sigma_A^Z&=\frac{G_F^2\mu_A^2}{8\pi}
 \left[N-(1-4s_W^2)Z\right]^2,\\
 \sigma_n^Z&=1.86\times10^{-39}\ {\rm cm^2}.
\end{align}
Using the weak-vector conversion specified by LZ, this value lies close to the
upper edge of the two-sided interval at $\delta_0=0.35$ MeV in their
\WY{Fig.~S7}.  A dedicated likelihood with the fixed proton/neutron couplings is
needed before assigning a Higgsino significance~\cite{LZ:2026highE}.

\paragraph{\WY{SUSY spectrum indicated by the event}}
Integrating out the bino and wino gives~\cite{Nagata:2014wma}
\begin{equation}
 \delta_0\simeq m_Z^2\left(\frac{s_W^2}{M_1}+\frac{c_W^2}{M_2}\right).
 \label{eq:splitting}
\end{equation}
For equal, aligned electroweak gaugino masses, $\delta_0=0.35$ MeV requires
$M_1=M_2=2.4\times10^7$ GeV.  This scale follows from a SUSY-breaking spurion
${\cal Z}$ through~\cite{Martin:1997ns}
\begin{align}
 \int d^2\theta\,\frac{c_a{\cal Z}}{4\mpl}{\cal W}^{a\alpha}{\cal W}^a_\alpha,
 &\qquad M_a=c_a\frac{F_{\cal Z}}{\mpl},\\
 \int d^4\theta\,\frac{c^f_{ij}{\cal Z}^\dagger{\cal Z}}{\mpl^2}
 Q_i^\dagger Q_j,
 &\qquad (m_{\widetilde f}^2)_{ij}=c^f_{ij}
 \frac{|F_{\cal Z}|^2}{\mpl^2}.
\end{align}
\WY{Because the spurion is
uncharged, sfermion trilinear $A$-terms are also allowed. Such terms can induce
$CP$ and flavor violation in low-energy SUSY}.
\WY{Analogous couplings of $H_u^\dagger H_u$ and $H_d^\dagger H_d$ to the
SUSY-breaking sector generate the Higgs soft masses, whereas operators
proportional to $H_uH_d$ are forbidden by the Peccei--Quinn (PQ) symmetry.}

Heavy sfermions suppress flavor violation and one-loop electric dipole moments, while heavy gauginos suppress the remaining electroweakino phases~\cite{Gabbiani:1996hi,Nagata:2014wma}.  The electroweak hierarchy is, however, fine-tuned as various SUSY model. Of course no acelarator limit can constrain the scenario. 

\WY{I impose} a PQ symmetry and use the Kim--Nilles
operator~\cite{Kim:1983dt}
\begin{equation}
\laq{axion}
 W\supset\lambda_\mu\frac{P^2}{\mpl}H_uH_d,
 \qquad
\vev{P}\simeq5.1\times10^{10}\lambda_\mu^{-1/2}\ {\rm GeV},
\end{equation}
which gives $|\mu|=1.08$ TeV and a QCD axion solving the strong-$CP$
problem~\cite{Peccei:1977hh}.  The axion may be subdominant dark matter.  
\WY{Supergravity effects then generate}
\beq
B_\mu = m_{3/2} \mu.
\eeq

  \WY{I perform} the usual
high-scale-SUSY fine-tuning of the Higgs mass matrix~\cite{Jeong:2011sg},
\begin{equation}
 {\cal M}_H^2=
 \begin{pmatrix}
 m_{H_u}^2+|\mu|^2 & -B_\mu\\
 -B_\mu^* & m_{H_d}^2+|\mu|^2
 \end{pmatrix},
 \label{eq:higgs-mass-matrix}
\end{equation}
so that one eigenstate remains at the electroweak scale.  The orthogonal
doublet has
\begin{equation}
 m_A^2=m_{H_u}^2+m_{H_d}^2+2|\mu|^2
       =\frac{2|B_\mu|}{\sin2\beta}.
 \label{eq:heavy-higgs-mass}
\end{equation}
It is generically of the Higgs-sector soft scale.  A heavy-Higgs mass below
the common sfermion scale can be accommodated by nonuniversal
gravity-mediated coefficients without introducing a separate
mediation mechanism~\cite{Martin:1997ns,Yamaguchi:2016oqz}.

\paragraph{Higgs mass and electroweak symmetry breaking}

\begin{figure}[t]
 \centering
 \includegraphics[width=\columnwidth]{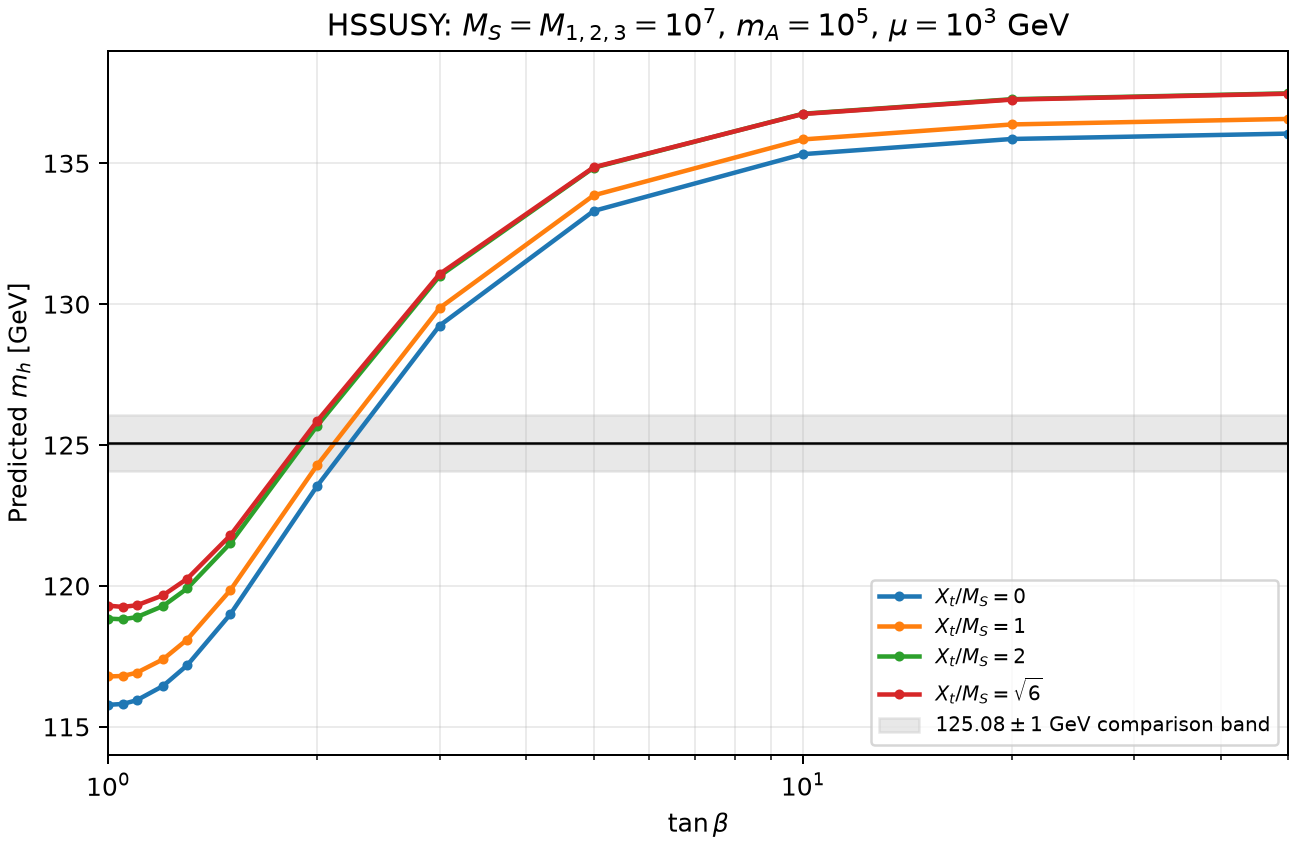}
 \caption{HSSUSY prediction for the light Higgs mass.  The gray band denotes
 $m_h=125.08\pm1\GEV$.}
 \label{fig:hssusy-higgs}
\end{figure}

\WY{I check} the Higgs mass with HSSUSY in FlexibleSUSY~\cite{Athron:2017fvs}.
For illustration, \WY{I take} $M_S=M_1=M_2=M_3=10^7\GEV$,
$m_A=10^5\GEV$, and $\mu=1\TEV$.
As shown in Fig.~\ref{fig:hssusy-higgs}, the observed Higgs mass is obtained
for $\tan\beta\simeq1.9$--$2.2$, \WY{depending on stop mixing}~\cite{Bagnaschi:2014rsa}.  For example,
$X_t=0$ gives $m_h\approx 125\GEV$ at $\tan\beta=2.20$.  The spectrum is therefore
consistent with the measured Higgs mass.

\begin{table*}[t]
 \caption{Illustrative spectrum.  The heavy-Higgs, sfermion, and gluino
 entries are threshold parameters to be refined by electroweak matching and
 the gauge-unification scan.}
 \label{tab:spectrum}
 \begin{ruledtabular}
 \begin{tabular}{lll}
 State or parameter & Illustrative value & Origin \\
 \hline
 $h$ & $125\ {\rm GeV}$ & tuned light eigenstate \\
 $\widetilde H^0_{1,2}$ & $1.08\ {\rm TeV}$ & thermal Higgsino DM \\
 $m_{\widetilde H^0_2}-m_{\widetilde H^0_1}$
   & $0.28$--$0.35\ {\rm MeV}$ & LZ recoil kinematics \\
 $m_{\widetilde H^\pm}-m_{\widetilde H^0_1}$
   & $\simeq0.35\ {\rm GeV}$ & electroweak loop \\
 $H,A,H^\pm$ & $m_A\sim10^5$--$10^7\ {\rm GeV}$
   & Higgs-sector coefficients \\
 $\widetilde Q,\widetilde U,\widetilde D,\widetilde L,\widetilde E$
   & ${\cal O}(10^7)\ {\rm GeV}$ & gravity mediation; may be split \\
$\widetilde B,\widetilde W$ & ${\cal O}(10^7)\ {\rm GeV}$
   & neutral-Higgsino splitting \\
$\widetilde g$ & ${\cal O}(10^8)\ {\rm GeV}$ & gauge unification; may vary \\
$a$ & $m_a\sim10^{-4}\ {\rm eV}$ & QCD axion \\
 \end{tabular}
 \end{ruledtabular}
\end{table*}

\section{Unification, proton decay, and the thermal history}

\begingroup
I evolve the gauge couplings at two loops, together with the one-loop top
Yukawa coupling, using sharp step thresholds.
\endgroup

\begin{figure*}[t]
 \centering
 \includegraphics[width=0.94\textwidth]{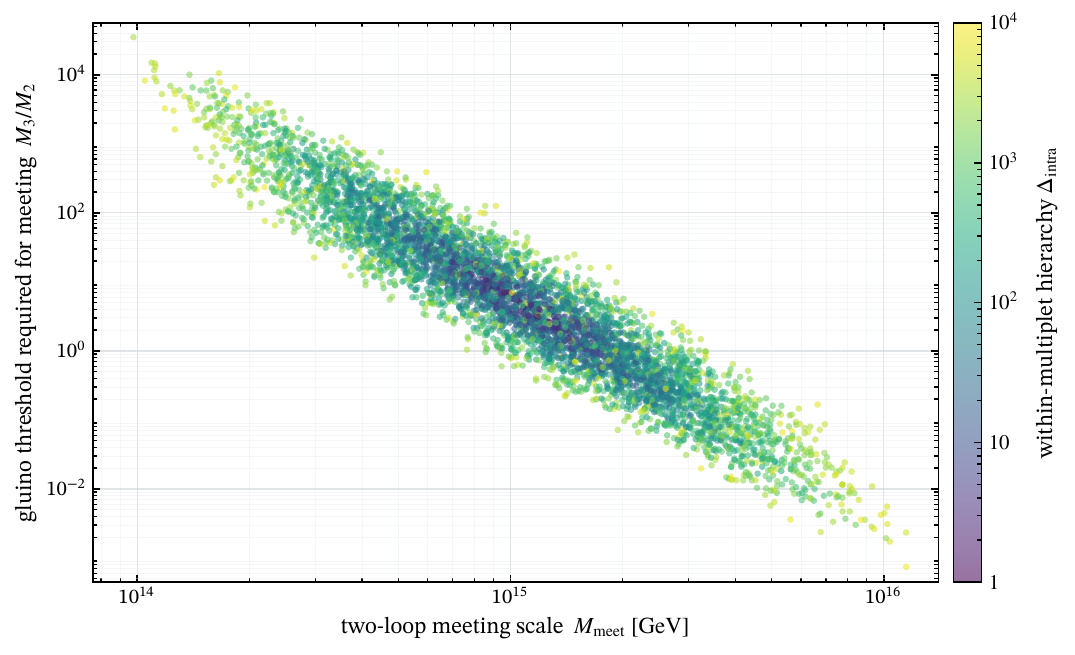}
 \caption{{Broad sfermion threshold scan.  The five
 generation-degenerate masses $m_{\widetilde Q}$, $m_{\widetilde U}$,
 $m_{\widetilde D}$, $m_{\widetilde L}$, and $m_{\widetilde E}$ are sampled
 independently and logarithmically over $10^5$--$10^9$ GeV, while the fixed
 inputs are given in the text.  At each point, $M_3$ is determined
 numerically so that the three gauge couplings meet.  The color denotes
 $\Delta_{\rm intra}\equiv\max\{\max(m_{\widetilde Q},m_{\widetilde U},
 m_{\widetilde E})/\min(m_{\widetilde Q},m_{\widetilde U},m_{\widetilde E}),
 \max(m_{\widetilde D},m_{\widetilde L})/\min(m_{\widetilde D},m_{\widetilde L})\}$.}}
 \label{fig:split-sfermion-unification}
\end{figure*}

\begingroup
Figure~\ref{fig:split-sfermion-unification} shows scatter plots with 5000 random fluctuations of sfermion masses. In particular, we can have large splitting sfermion mass spectra, e.g.,~\cite{Yamaguchi:2016oqz}.  For instance, among
the sampled spectra, 168 have $M_{\rm meet}\geq5\times10^{15}$ GeV.  Every
one has the mass slitting $\Delta_{\rm intra}\geq253$. Thus the high-scale tail is not produced
by an order-one threshold perturbation; it also selects a correlated lighter
gluino threshold, as shown on the vertical axis. The scan establishes a
possible threshold pattern.
\endgroup

\paragraph{Dimension-five and dimension-six proton decay}
\begingroup
Two proton-decay mechanisms should be distinguished.  Colored-Higgsino
exchange generates the supersymmetric dimension-five operators
$QQQL/M_{H_C}$ and $u^cu^cd^ce^c/M_{H_C}$, whose wino or Higgsino dressing
predominantly gives $p\to K^+\bar\nu$.  For the heavy-sfermion and small-
$\tan\beta$ spectrum considered here this contribution is strongly
decoupled; we assume in addition that the PQ/flavor structure suppresses
independent Planck-suppressed dimension-five operators.  Under these
assumptions dimension-five decay can satisfy the present bounds, although a
specific GUT completion still requires a Wilson-coefficient calculation
~\cite{Hisano:2013revive,Nagata:2014proton,Dine:2014proton,SuperK:2014pknu}.

The relevant unification-scale test is then dimension-six $X,Y$ gauge-boson
exchange, with the canonical estimate
\begin{equation}
 \frac{\tau(p\to e^+\pi^0)}{B}
 \simeq 10^{34}\ {\rm yr}
 \left(\frac{M_X}{5\times 10^{15}\ {\rm GeV}}\right)^4
 \left(\frac{0.033}{\alpha_U}\right)^2 ,
 \label{eq:proton-d6}
\end{equation}
up to order-one hadronic, renormalization, and GUT-flavor factors
~\cite{Hisano:2022proton,Ellis:2020proton}.  The Super-Kamiokande limit
$\tau/B(p\to e^+\pi^0)>2.4\times10^{34}$ yr at 90\% C.L. corresponds
canonically to $M_X\gtrsim6\times10^{15}$ GeV for $\alpha_U\simeq0.033$
~\cite{Super-Kamiokande:2020wjk}.  We stress that the RGE crossing
$M_{\rm meet}$ is not generally the physical gauge-boson mass $M_X$:
already in minimal supersymmetric $SU(5)$, threshold matching constrains
$(M_X^2M_\Sigma)^{1/3}$ rather than $M_X$ alone~\cite{Hisano:2013gut}.
Thus points with $M_{\rm meet}\sim(6$--$10)\times10^{15}$ GeV are near
present and next-generation dimension-six sensitivity only if GUT
thresholds place $M_X$ near $M_{\rm meet}$; an explicit GUT spectrum is
required for a lifetime prediction.
\endgroup

\paragraph{Cosmological consistency}
Consider a modulus or gravitino-like state $X$ with
$\Gamma_X=c_Xm_X^3/(2\pi\mpl^2)$~\cite{Nakamura:2006uc,Asaka:2006bv}.  Its decay temperature is
\begin{equation}
 T_D\simeq100\ {\rm GeV}\,c_X^{1/2}
 \left(\frac{200}{g_*}\right)^{1/4}
 \left(\frac{m_X}{10^8\ {\rm GeV}}\right)^{3/2}.
\end{equation}
This is safely before nucleosynthesis and can  preserve a
thermal Higgsino abundance: $T_f\simeq m_{\widetilde H}/20\simeq50$ GeV.
Thus the conventional late-decay gravitino and modulus problems can be
avoided.

\section{Discussions}
The thermal Higgsino line signal is close to the new H.E.S.S. \WY{Galactic Center}
sensitivity but is not excluded for the baseline Einasto profile~\cite{Beneke:2019gtg,HESS:2026ila}.
Thus the further search of the indirect detection may also  probe our scenario.

\WY{The QCD axion that solves the strong-$CP$ problem can provide a
subdominant dark-matter component for $\l_\mu=\O(1)$ [see \Eq{axion}] through
the misalignment mechanism~\cite{Preskill:1982cy,Abbott:1982af,Dine:1982ah}}, \WY{assuming that the PQ
symmetry is not restored after inflation. For $\l_\mu<\O(1)$ while matching the TeV scale, the PQ scale is
larger and the axion can constitute a non-negligible fraction of dark matter.
In standard cosmology this requires a slightly lighter Higgsino so that its
thermal relic abundance does not saturate the observed dark-matter density
~\cite{Nagata:2014wma}.}

\WY{Then the
annihilation flux scales as the inverse of the annihilation crosssection, which gets enhanced by considering the lighter Higgsino.  Given the uncertainty in the direct detection side, and the slight change of mass splitting is allowed for event rate, this can alleviate the constraint from the indirect detection experiments.}

In the minimal scenario, this model predicts the axion with the small photon coupling via an accidental cancellation with $E/N=2$ due to the Higgsino contribution~\cite{GrillidiCortona:2015jxo}, 
\beq g_{a\gamma\gamma}
=
\frac{\alpha}{2\pi f_a}
\left(\frac EN-1.92(4)\right)
=
\frac{\alpha}{2\pi f_a}\,[0.08(4)].\eeq
Probing this super-invisible axion, together with $p\to e^+\pi$ and Higgsino dark matter, would provide a smoking-gun signature of this scenario. It may therefore be important to extend the target sensitivity of axion searches to this range.

\section*{acknowledgement}
This work is supported by JSPS KAKENHI Grant Nos. 22K14029 (W.Y.),
23K22486 (W.Y.), and 26K00695 (W.Y.).  W.Y. is also supported by the Selective
Research Fund and the Incentive Research Fund of Tokyo Metropolitan University.
 Codex and ChatGPT were used for reference searches, calculations, and writing assistance based on my original idea. I take full responsibility for the content.

\bibliography{inelastic_thermal_higgsino_refs}

\end{document}